\documentclass[10pt,preprint,aps,prc,showpacs]{revtex4-1}
\usepackage{graphicx,epstopdf,multirow,amsmath,rotating,enumitem,lmodern}

\begin{document}


    \author{Deepak Kumar}
    \author{Moumita Maiti}
    \email{moumifph@iitr.ac.in, moumifph@gmail.com}
	\affiliation{Department of Physics, Indian Institute of Technology Roorkee, Roorkee--247667, Uttarakhand, India}
	\author{Susanta Lahiri}
	\affiliation{Chemical Sciences Division, Saha Institute of Nuclear Physics, 1/AF Bidhannagar, Kolkata--700064, India}
	\title{$^{\text{7}}$Li-induced reaction on $^{\text{nat}}$Mo: A study of complete versus incomplete fusion}
	\date{\today}

		\begin{abstract}
		\begin{description}
\item[Background]
Several investigations on the complete-incomplete fusion (CF-ICF) dynamics of $\alpha$-cluster well-bound nuclei has been contemplated above the Coulomb barrier ($\sim$ 4--7 MeV/nucleon) in recent years. It is therefore expected to observe significant ICF over CF in the reactions induced by a weakly bound $\alpha$-cluster nucleus at slightly above the barrier.
\item[Purpose]
Study of the CF-ICF dynamics by measuring the populated residues in the weakly bound $^{\text{7}}$Li+$^{\text{nat}}$Mo system at energies slightly above the Coulomb barrier to well above it.  
\item[Method] 
In order to investigate CF-ICF in the loosely bound system, $^{\text{7}}$Li beam was bombarded on the $^{\text{nat}}$Mo foils, separated by the aluminium (Al) catcher foils  alternatively, within $\sim$ 3--6.5 MeV/nucleon. Evaporation residues produced in each foil were identified by the off-line $\gamma$-ray spectrometry. Measured cross section data of the residues were compared with the theoretical model calculations based on the equilibrium (EQ) and preequilibrium (PEQ) reaction mechanisms. 

\item[Results] 
The experimental cross section of $^{\text{101m,100,99m,97}}$Rh, $^{\text{97,95}}$Ru, $^{\text{99m,96,95,94,93m+g}}$Tc and $^{\text{93m}}$Mo residues measured at various projectile energies were satisfactorily reproduced by the simplified coupled channel approach in comparison to single barrier penetration model calculation. Significant cross section enhancement in the $\alpha$-emitting channels was observed compared to EQ and PEQ model calculations throughout observed energy region. The ICF process over CF was analyzed by comparing with {\fontsize{9}{10}\selectfont EMPIRE}. The increment of the incomplete fusion fraction was observed with increasing projectile energies.
			
\item[Conclusions] 
Theoretical model calculations reveal that compound reaction mechanism is the major contributor to the production of residues in $^{\text{7}}$Li+$^{\text{nat}}$Mo reaction. Theoretical evaluations substantiate the contribution of ICF over the CF in $\alpha$-emitting channels. {\fontsize{9}{10}\selectfont EMPIRE} estimations shed light on its predictive capability of cross sections of the residues from the heavy-ion induced reactions.
     \end{description}
	\end{abstract}
   
  \maketitle

	
	\section{\label{s1}Introduction}
	
	Enormous efforts have been made to study fusion-like events, viz., complete fusion (CF), incomplete fusion (ICF) and direct processes, such as breakup or transfer in many medium-energy weakly bound heavy-ion (A $\le$ 20) induced reactions on intermediate/heavy mass nuclei around the Coulomb barrier to well above it \cite{canto15}. A consistent systematic comparison of experimental results reveals the enhancement of CF cross section at sub-barrier energies and suppression above it leading to ICF contribution in loosely bound projectiles. However, the total fusion (TF) cross section was satisfied by the coupled channel (CC) calculations indicating an insignificant influence of continuum or the transfer channel on TF, whereas, enhancement was observed below the barrier in heavy mass nuclei \cite{canto15, fang15, gomes06, gomes04}.
	Moreover, significant ICF over the CF above the barrier $\sim$ 4--10 MeV/nucleon in well-bound $\alpha$-cluster ion induced reactions has also been observed in the $\alpha$-emitting channels over the past few years \cite{yadav12, sharma14, mukherjee06, hkumar17, shuaib16,unnati09}. More experimental investigation is therefore necessary to resolve discrepancies in the dependence of CF-ICF on projectile types at the low incident energies, particularly, for weakly bound nuclei in which breakup yield is more probable in the nuclear force field.

	In a CF process, projectiles having low impact parameter or angular momentum ($l$) smaller than critical angular momentum ($l_{crit}$) where the attractive nature of potential disappears at the point of contact of interacting nuclei, coalesce completely with the target by overcoming the nuclear interaction and form a completely fused excited composite system. In a fusion process following breakup, one possibility is that only a fragmented cluster of projectile blend with the target, known as incomplete or partial fusion, forming composite system with smaller charge, mass, excitation energy, momentum, and angular momentum compared to the CF process. Remnant part of the projectile (another cluster) flies away towards the forward direction with velocities near that of the projectile. Another possibility is the sequential merging of the fragments into the target which could not be experimentally distinguished from the CF process. Experimentally, one can usually measure total fusion (CF + ICF) cross-section, except some special cases, for instance, a heavy mass nuclei where neutron emission (apart from fission) is only dominant channel and emission of charge particles are usually hindered due to large Coulomb barrier \cite{parker84}. For the weakly bound projectiles having low breakup thresholds, fusion-like processes can be influenced either by static effects as the shape of the target-projectile potential may affected by large diffuseness in density distribution or by dynamical effects due to coupling to bound states as well as to the continuum. Moreover, ICF processes in weakly bound particle induced reactions may also be favoured due to considerable breakup yield. Thus CF and ICF are two distinct and independent processes, having the characteristic velocity distribution of populated residues, utilized in the recoil range distribution techniques for CF-ICF study \cite{parker84, parker89, tomar94, bkumar98, tomar98}. However, it it worthy to mention at this point that contribution of ICF process may include also the direct processes, which would eventually lead to the same effect as ICF.

		About half a century ago, forward peaked high energetic $\alpha$-particles were observed in the intermediate energy heavy-ion induced reactions leading to the considerable partial fusion due to merging of remaining part of a projectile with the target, termed as massive transfer reactions \cite{britt61, galin74, inamura77, zolnowski78}. In the eighties and nineties, a number of light heavy-ions (such as $^{\text{12}}$C, $^{\text{16}}$O, $^{\text{20}}$Ne etc.) induced reactions on heavy nuclei was performed to investigate CF-ICF using particle-gamma coincidence techniques and recoil range distribution techniques \cite{barker80, lunardon99, parker84, parker89, hogan85, tomar94, bkumar98, tomar98, chakrabarty00}. It is observed that the articles \cite{inamura77, zolnowski78, barker80, lunardon99} are consistent with the sum rule model \cite{wilczynska79} and \cite{tomar94, bkumar98, tomar98, chakrabarty00} with the breakup fusion model \cite{udagawa80}, respectively.  Besides, the peripheral nature of CF-ICF was elaborated by Geoffroy {\it et al.}, Trautmann {\it et al.}, Inamura {\it et al.}, Zolnowski {\it et al.}, and Oeschler {\it et al.} \cite{geoffroy79, trautmann84, zolnowski78, inamura77, oeschler83}. The dependence of the localization of the entrance channel angular momentum ($l$) window on deformation of target nucleus was well described in the review of Gerschel {\it et al.} \cite{gerschel82}.
		However, ICF was also observed at the low projectile energies near barrier which mainly involve $l$ $\le$ $l_{crit}$ reported by Tricoire et al, Utsunomiya {\it et al.}, Tserruya {\it et al.} \cite{tricoire86, utsunomiya81, tserruya88}.
		
		From the past decade, several efforts were made to study the effects of various entrance channel parameters, like projectile energy and its structure, angular momentum ($l$) window for ICF, deformation of interacting nuclei, mass asymmetry, $\alpha$-Q values (Q$_\alpha$) etc.\ on the ICF process by the recoil range distribution technique/recoil-catcher activation technique \cite{yadav12, sharma14, mukherjee06, hkumar17, shuaib16, unnati09}.
		An increment in the breakup probability of a projectile with the increasing bombarding energy, more ICF in $\alpha$-cluster nuclei than others and more ICF for larger entrance channel mass asymmetry corresponding to the same relative velocity, significant ICF contribution for $l \le l_{crit}$ contrary to the sum rule model, has been observed in $^{\text{12,13}}$C induced reactions on $^{\text{181}}$Ta, $^{\text{159}}$Tb, $^{\text{169}}$Tm, $^{\text{175}}$Lu \cite{yadav12, sharma14, mukherjee06, hkumar17}.
		Incomplete fusion fraction (F$_{\text{ICF}}$), a measure of the relative strength of ICF to the total fusion, was observed independent of target charge (Z$_t$) \cite{rafiei10, gasques09}; however, contrary to that, in Refs \cite{hinde02, gomes11, shuaib16} its nature is found almost proportional to charge Z$_t$.
		
		Breakup probability in loosely bound nuclei ($^{\text{6,7}}$Li, $^{\text{9}}$Be etc.) is expected to be large in comparison to well-bound nuclei \cite{rafiei10, gasques09}. Therefore, understanding of the dynamics of CF and ICF near the barrier became more important in the reactions induced by weakly bound projectiles. Recently, suppression and enhancement of the CF cross section at the energies above and below the barrier were observed in $^{\text{10,11}}$B+$^{\text{209}}$Bi; $^{\text{6,7}}$Li+$^{\text{197}}$Au,$^{\text{209}}$Bi; $^{\text{7}}$Li+$^{\text{65}}$Cu; $^{\text{9}}$Be+$^{\text{144}}$Sm,$^{\text{208}}$Pb reactions in comparison to coupled channel predictions that do not consider the couplings of projectile continuum \cite{gasques09,palshetkar14, shrivastava06, hinde02, gomes11, esbensen10}. Badran {\it et al.} \cite{badran01} investigated the CF and ICF from the produced residues in $^{\text{7}}$Li + $^{\text{56}}$Fe by observing inclusive light particle spectra along with recoil range distribution of residual products. A speculation was also drawn for ICF in $^{\text{7}}$Li+$^{\text{93}}$Nb reaction \cite{dkumar16}. Moreover, our group also have studied $^{\text{7}}$Li, $^{\text{11}}$B, $^{\text{12}}$C induced reactions on the targets around A$\sim$ 90 to produce $^{\text{97}}$Ru (2.83 d) and $^{\text{101m}}$Rh (4.34 d) radionuclides for the application purposes \cite{maiti11,maiti13,maiti15,dkumar17,dm17,dm2prc17}.
		
		In the present article, study of CF-ICF processes for $^{\text{7}}$Li-induced reaction on the natural molybdenum ($^{\text{nat}}$Mo) target have been carried out in the $\sim$ 3--6.5 MeV/nucleon energy range. The excitation function of $^{\text{101m,100,99m,97}}$Rh, $^{\text{97,95}}$Ru, $^{\text{99m,96,95,94,93m+g}}$Tc and $^{\text{93m}}$Mo radionuclide are analyzed by comparing with the theoretical model calculations from {\fontsize{9}{10}\selectfont PACE}4 and {\fontsize{9}{9}\selectfont EMPIRE}3.2. Enhancement of cross sections at the $\alpha$-emitting channels was also observed and attributed to ICF over CF process.

		The experimental procedure and a comparative study of the nuclear model calculations are presented in section \ref{s2}. Section \ref{s3} discusses the results of the present study and \ref{s4} summarizes the report.

		
    \section{\label{s2}Experimental Details}
    \subsection{Irradiation}
		Spectroscopically pure (99.99\%) natural molybdenum ($^{\text{nat}}$Mo) target, procured from Alfa Aesar was bombarded by the $^{7}$Li beam delivered by the BARC-TIFR Pelletron facility Mumbai, India. Self-supporting $^{\text{nat}}$Mo foils  and aluminum (Al) foils were prepared by proper machine rolling. In each stack arrangement, 3 sets of Mo--Al pair having a thickness between 2.2--3 and 2 mg/cm$^\text{2}$, respectively, were taken and six of such stacks were exposed to beam with overlapping energy to encompass the $\sim$ 20--45 MeV energy range. The Al foils, which also served as an energy degrader, were alternately placed behind the target for the complete collection of recoiled residues produced during the bombardment of beam on the $^{\text{nat}}$Mo foils. The Mo--Al foils were mounted on the annular Al holders having an inner and outer diameter 12 and 22 mm, respectively. The irradiation time was judged on the basis of beam current and the half-lives of the populated residues in the $^{\text{7}}$Li + $^{\text{nat}}$Mo. The $^\text{7}$Li energy at each target is taken as the average of the entrance and exit beam energy, calculated by using stopping and range of ions in matter ({\fontsize{9}{10}\selectfont SRIM}) software. The total charge of each irradiation was collected by an electron suppressed Faraday cup fixed at the rear side of the target assembly and was measured by using a digital current integrator. Beam flux was kept almost constant during the experiment. 
		
	\subsection{Identification of residues }
	After the end of bombardment (EOB), the activity of populated radionuclides in each target matrix was determined using the $\gamma$-ray spectrometry employing an n-type HPGe detector having an energy resolution of 2.1 keV at 1.33 MeV. Several time resolved $\gamma$-ray spectra were acquired for a sufficiently long time period and analyzed using a digital spectrum analyser (DSA) and GENIE-2K software.
	 The energy and efficiency calibration of the detector was performed by using standard sources: $^{\text{152}}$Eu (13.53 a), $^{\text{137}}$Cs (30.08 a), $^{\text{60}}$Co (5.27 a), $^{\text{133}}$Ba (10.51 a) of known strength. Each Mo foil along with an Al foil were collectively analyzed using a proper geometry so that the dead time of the detector should be less than 5\% for all measurements. The radionuclides were identified by their characteristics $\gamma$-ray energies and corresponding decay curve. Nuclear spectroscopic data of the residual radionuclides are listed in Table \ref{t1} \cite{nndc}.
	Details of the activity calculation, cross-section estimation and corresponding uncertainties are described in our previous reports \cite{ms11,mm11,mm10}. The main sources of errors in the cross-section measurement are fluctuation in beam flux ($\sim$ 6--7\%), non-uniformity of foils and in measuring its thickness ($\sim$ 2--3\%), statistical error in $\gamma$-ray counting ($\sim$ 1--7\%) and efficiency calibration of the detector ($\sim$ 0.1--1\%). The measured data are reported in this article up to 95\% confidence level. The error in the incident projectile energies includes the error in the ({\fontsize{9}{10}\selectfont SRIM}) calculation and determination of the target thickness.

	\subsection{Distinctive features of {\fontsize{9}{10}\selectfont PACE}4 and {\fontsize{9}{10}\selectfont EMPIRE}3.2}
	
	Contribution of CF and ICF from the measured cross section data has been accomplished with the help of theoretical model codes: {\fontsize{9}{10}\selectfont EMPIRE}3.2 \cite{herman07} and {\fontsize{9}{10}\selectfont PACE}4 \cite{gavron80}. {\fontsize{9}{10}\selectfont EMPIRE} takes into account all three major reaction formalisms: Direct, PEQ, and EQ processes, while {\fontsize{9}{10}\selectfont PACE} considers only statistically equilibrated compound nuclear system (EQ). A brief comparative analysis of both the codes has been presented below:

	\begin{itemize}[leftmargin=*]
	\item Emission of direct particles is estimated in {\fontsize{9}{10}\selectfont EMPIRE} by coupled channel approach or distorted wave Born approximation (DWBA) method \cite{raynal,raynal72}. There is no such provision to account the direct processes in {\fontsize{9}{10}\selectfont PACE}.
  \item Hauser-Feshbach formalism was adopted for evaporation of particles by both, {\fontsize{9}{10}\selectfont EMPIRE} and {\fontsize{9}{10}\selectfont PACE}; however {\fontsize{9}{10}\selectfont EMPIRE} also includes the width fluctuation correction factor to account correlation between entrance and exit channel inelastic scattering.
  \item {\fontsize{9}{10}\selectfont EMPIRE} utilizes an exciton model for PEQ emission of particles in heavy-ion induced reactions employing capability of cluster emission in terms of the Iwamotto Harada model. Multistep direct (MSD), multistep compound (MSC), hybrid Monte Carlo simulation approach can also be used for a light particle induced reactions; however, {\fontsize{9}{10}\selectfont PACE} does not consider PEQ emissions.	
\item In {\fontsize{9}{10}\selectfont EMPIRE}, simplified coupled channels approach \cite{dasso87} or distributed fusion barrier model can be chosen for the transmission coefficient calculation, and subsequently complete fusion cross section. {\fontsize{9}{10}\selectfont PACE} adopts simple one dimensional tunnelling fusion barrier using Bass potential \cite{bass77}.
	\item Various level density options such as an enhanced generalized superfluid model (EGSM), generalized superfluid model (GSM), Gilbert-Cameron (GC) level density model, and microscopic combinatorial level density (HFBM) are provided in {\fontsize{9}{10}\selectfont EMPIRE}, whereas GC level density is the only option available in {\fontsize{9}{10}\selectfont PACE}.
	\item Energy-dependent Ignatyuk level density parameter is used in {\fontsize{9}{10}\selectfont EMPIRE}. In {\fontsize{9}{10}\selectfont PACE}, level density parameter, $a$ = A/k, where A, and k are the mass number and free adjustable parameter, respectively, is used.
	\item {\fontsize{9}{10}\selectfont EMPIRE} calculates isomeric state production cross section separately apart from total (sum of ground and isomeric state) cross section, while {\fontsize{9}{10}\selectfont PACE} estimates only total production cross section of a particular residue.
	\item Fission coefficient can be selected in {\fontsize{9}{10}\selectfont EMPIRE} depending on the type of projectiles: an optical model for fission induced by light particles and the Sierk model for heavy-ion induced fission. In {\fontsize{9}{10}\selectfont PACE}, fission is considered as a decay mode, and is calculated using modified rotating liquid drop fission barrier by A. J. Sierk \cite{sierk86}.
\end{itemize}

 In order to appreciate the predictability and effectiveness of both the codes for the heavy-ion induced reactions, theoretical cross sections obtained from {\fontsize{9}{10}\selectfont PACE} and {\fontsize{9}{10}\selectfont EMPIRE} are compared with the experimental data. Cross sections were estimated from all the stable isotopes of Mo ($^{92,94,95,96,97,98,100}$Mo) and were added according to their natural abundances. In the {\fontsize{9}{10}\selectfont PACE} calculation, level density parameter $a$ = A/10 was used and quantum mechanical tunnelling barrier was selected. In the {\fontsize{9}{10}\selectfont EMPIRE}, exciton model with 1.5 mean free path and EGSM level density were used, simplified coupled channel calculation was used for the estimation of fusion cross section.

	 
	\section{\label{s3}Results and discussion}
	
\begin{figure*} [t]
	 	\includegraphics[width=85mm]{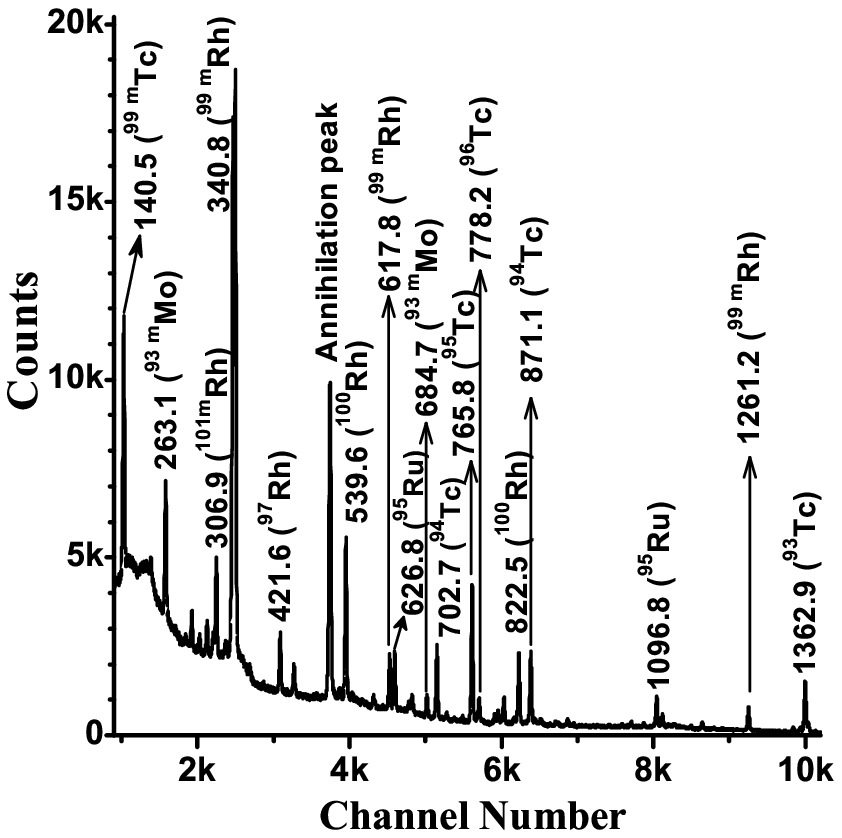}
	 	\caption{\label{fig1} A $\gamma$-ray spectrum of the $^{7}$Li+$^{\text{nat}}$Mo reaction at 42.8 MeV energy after 1.2 h of EOB (Characteristic $\gamma$-ray peaks are shown in keV).}
	 \end{figure*}

	A systematic identification and quantification of populated residues in the $^{\text{7}}$Li-induced reaction on $^{\text{nat}}$Mo target was performed in the 20--45 MeV energy range. A characteristic $\gamma$-ray spectrum of produced radionuclides at 42.8 MeV projectile energy collected after 1.2 hours of the EOB is portrayed in Fig \ref{fig1}. Accelerator produced $^{\text{101m,100,99m,97}}$Rh, $^{\text{97,95}}$Ru, $^\text{{99m,96,95,94,93m+g}}$Tc, and $^{\text{93m}}$Mo residues in the $^{\text{nat}}$Mo target matrix was detected and independent production of these radioisotopes are depicted in Fig \ref{fig2}--\ref{fig5}. CF and ICF cross section and variation of incomplete fusion fraction (F$_{\text{ICF}}$) with the projectile energies are presented in Fig \ref{fig6}. Experimental excitation functions are compared with the theoretical model calculation from {\fontsize{9}{10}\selectfont PACE} and {\fontsize{9}{10}\selectfont EMPIRE}. Measured data points are shown by symbols with an uncertainty and the theoretical calculations are shown by curves. The residues and the possible contributing reactions along with their threshold energy ($E_{th}$) are shown in Table \ref{t1}. Measured cross section data are presented in Tables \ref{t2} and \ref{t3}. The radionuclides produced through various reaction channels in the $^{\text{7}}$Li + $^{\text{nat}}$Mo system are discussed below:        
	
	\subsection{$xn$ and $pxn$ channels}
\begin{figure*} [t]
		 \includegraphics[width=180mm]{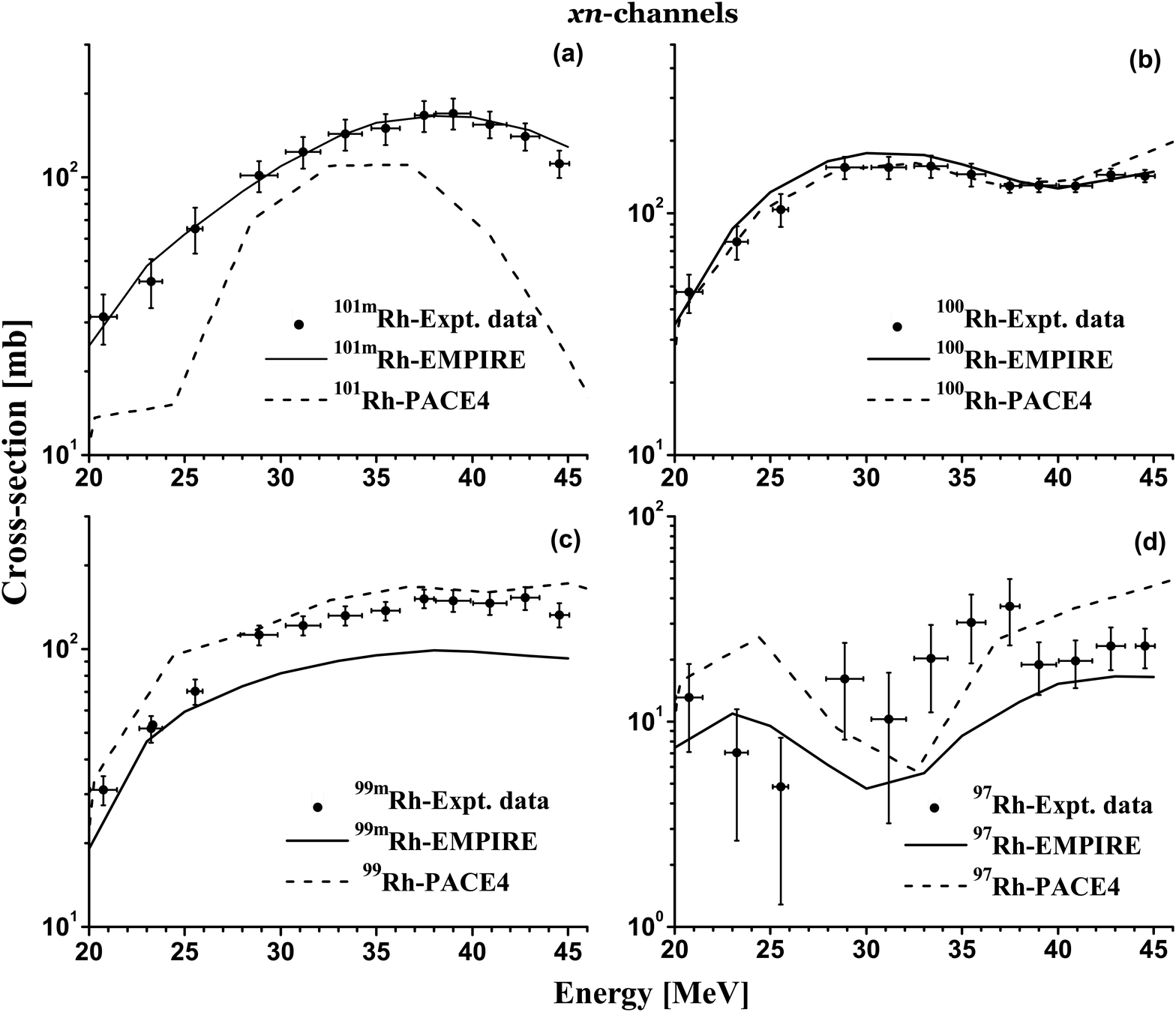}
		 \caption{\label{fig2} Excitation functions of the $^{\text{101m,100,99m,97}}$Rh radionuclide populated through $xn$-channels are compared with {\fontsize{9}{10}\selectfont EMPIRE} and {\fontsize{9}{10}\selectfont PACE} calculations.} 
		\end{figure*}

	Out of the ground (3.3 y) and isomeric (4.34 d) state of $^{\text{101}}$Rh, independent/direct production of $^{\text{101m}}$Rh was identified and it is shown in the Fig \ref{fig2}(a) along with the theoretical estimations. {\fontsize{9}{10}\selectfont EMPIRE} reproduces the experimental cross section fairly well; however {\fontsize{9}{10}\selectfont PACE} underpredicts them.  
	 The measured excitation function of ground state $^{\text{100}}$Rh, shown in the Fig \ref{fig2}(b), is well explained by both the model calculations. On the basis of {\fontsize{9}{10}\selectfont PACE} and {\fontsize{9}{10}\selectfont EMPIRE} isotopic cross section calculations, it turns out that the production of $^{\text{101m}}$Rh and $^{\text{100}}$Rh are contributed mainly by the higher and intermediate isotopes of $^{\text{nat}}$Mo respectively, shown in Table \ref{t1}. 
	 The production cross sections of the isomeric state, $^{\text{99m}}$Rh (4.7 h,  EC + $\beta^+$ $\ge$ 99.84\%) via $xn$ channel is shown in Fig \ref{fig1}(c) along with model calculations. No sign of the production of $^{\text{99g}}$Rh (16.1 d, EC + $\beta^+$ (100\%)) was observed in this experiment. Though {\fontsize{9}{10}\selectfont EMPIRE} estimate is in agreement with the measured data in the low energy region, yet it underestimates the experimental cross section grossly; however, {\fontsize{9}{10}\selectfont PACE} shows better agreement with the data in the higher energy side. The probable contributory reactions that produced $^{\text{99m}}$Rh are listed in Table \ref{t1}. 

  Fig.\ \ref{fig2}(d) shows the measured and theoretical excitation functions for $^{\text{97}}$Rh. Overall, {\fontsize{9}{10}\selectfont EMPIRE} results reproduce the cross sections better in comparison to the {\fontsize{9}{10}\selectfont PACE}. It is worthy to mention that due to the short half-life and availability of only one unique characteristic $\gamma$-ray peak with low statistics of $^{\text{97}}$Rh, large uncertainties in the measured values are observed. 
  
\begin{figure*} [t]
	    	\includegraphics[width=180mm]{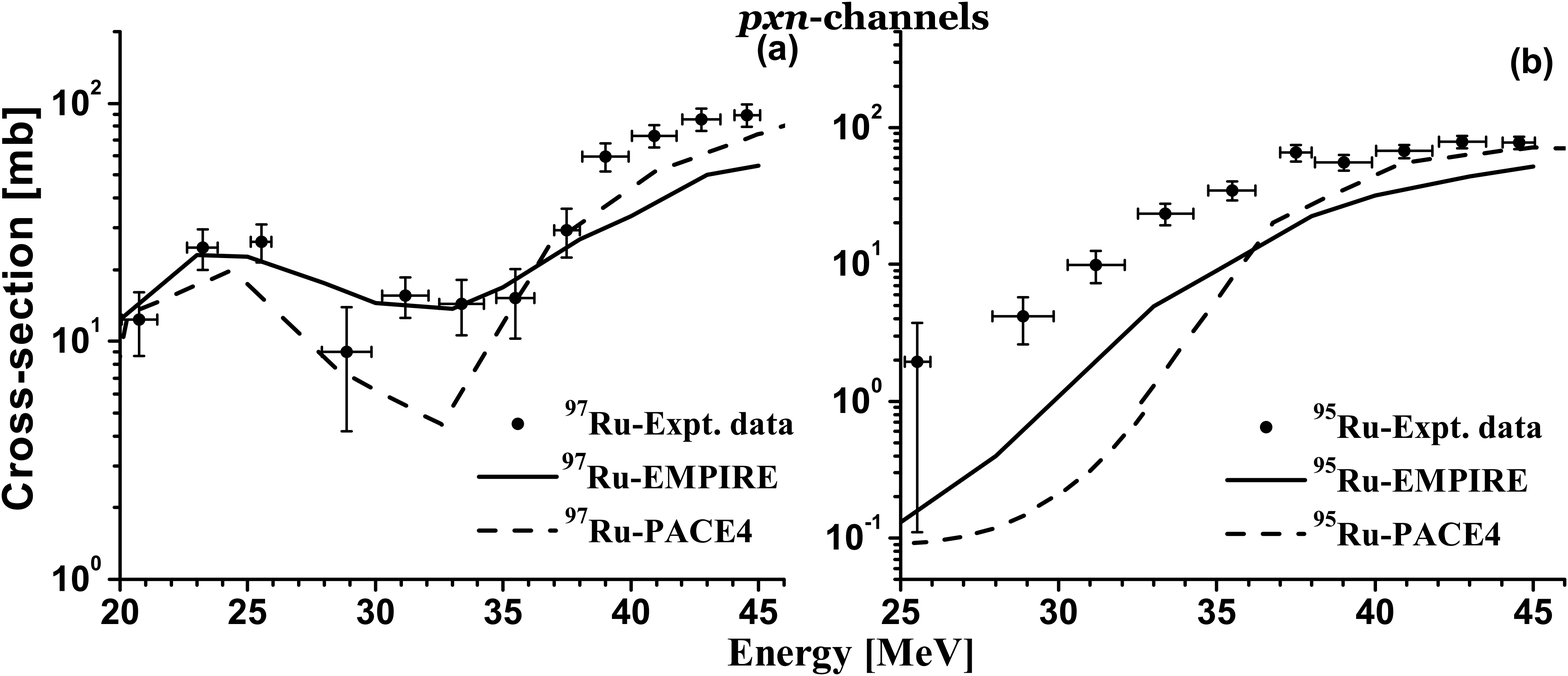}
	    	\caption{\label{fig3} Excitation functions of the $^{\text{97,95}}$Ru populated through $pxn$-channels are compared with {\fontsize{9}{10}\selectfont EMPIRE} and {\fontsize{9}{10}\selectfont PACE} calculations.}
	    \end{figure*}

\begin{figure*} [t]
	  	  	\includegraphics[width=180mm]{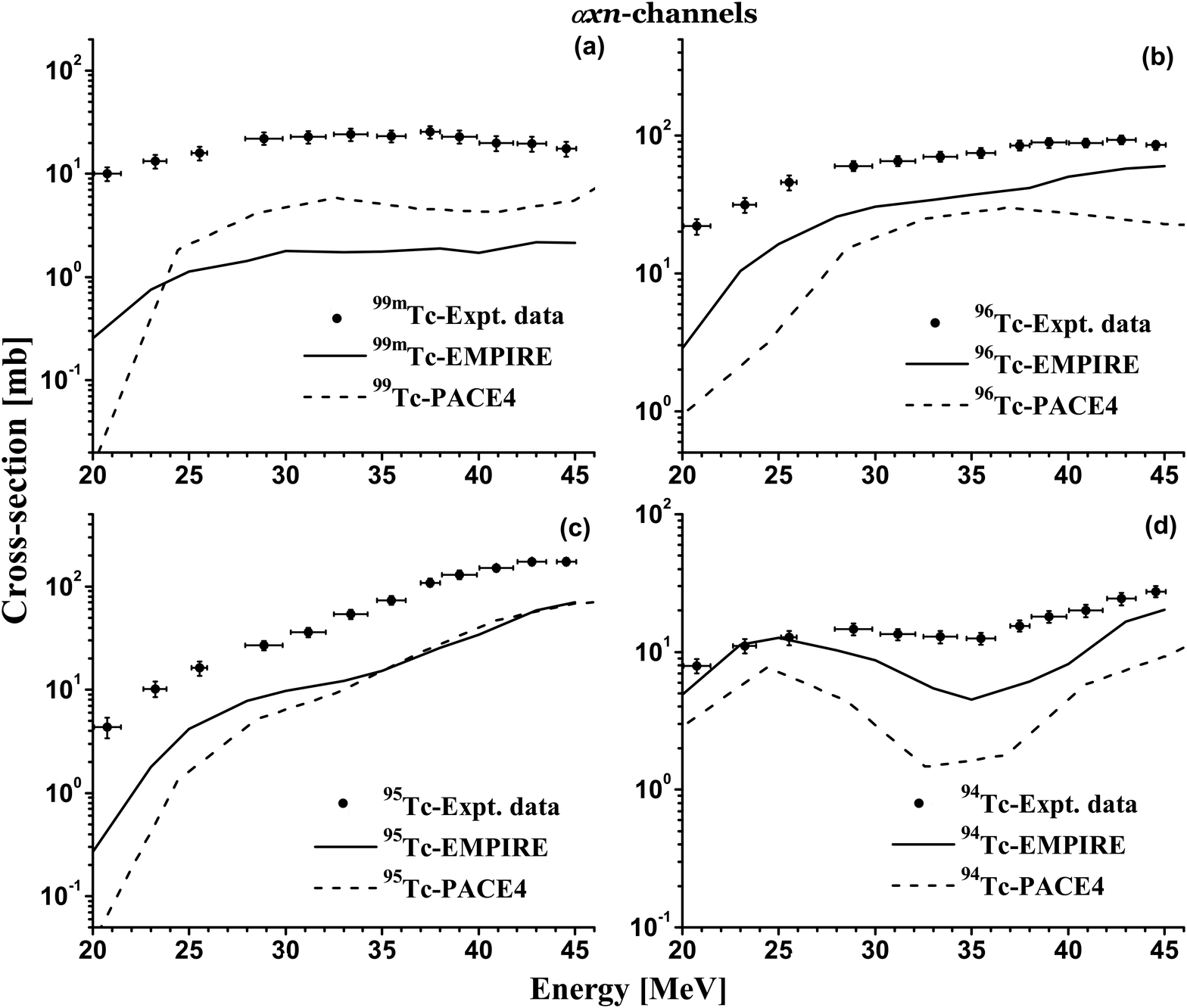}
	  	  	\caption{\label{fig4} Excitation functions of the $^{\text{99m,96,95,94}}$Tc residues are compared with {\fontsize{9}{10}\selectfont EMPIRE} and {\fontsize{9}{10}\selectfont PACE} calculations.}
	  	 \end{figure*}

\begin{figure*} [t]
	  	 	\includegraphics[width=180mm]{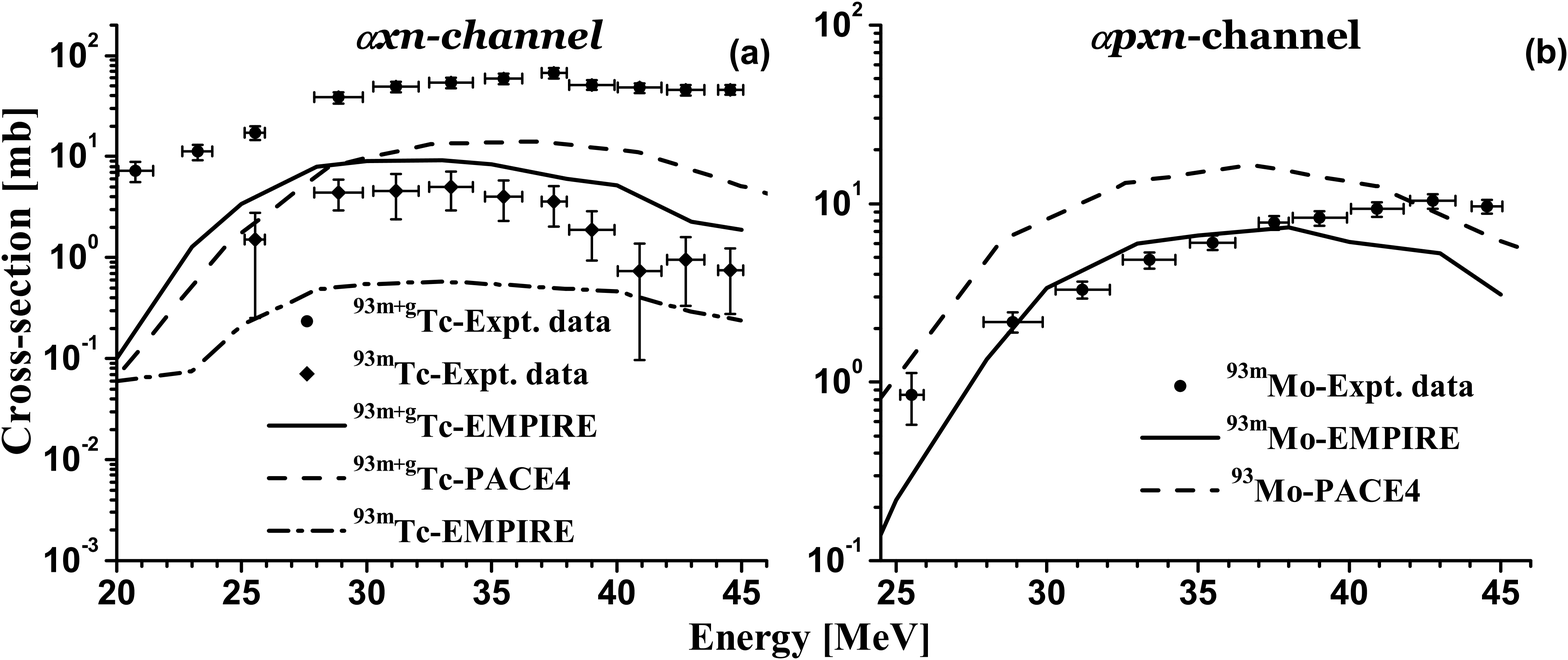}
	  	 	\caption{\label{fig5} Excitation functions of the $^{\text{93m+g}}$Tc, $^{\text{93m}}$Mo residues are compared with {\fontsize{9}{10}\selectfont EMPIRE} and {\fontsize{9}{10}\selectfont PACE} calculations.}
	  \end{figure*}

        The cross section values of $^{\text{97,95}}$Ru along with the theoretical curves are presented in Fig \ref{fig3}(a),(b). {\fontsize{9}{10}\selectfont EMPIRE} explains the experimental results of $^{\text{97}}$Ru (Fig \ref{fig3}(a)) up to 38 MeV incident energy and underpredicts above it. Fig \ref{fig3}(b) depicts that {\fontsize{9}{10}\selectfont EMPIRE} underestimates the measured data throughout estimated energy range; however, {\fontsize{9}{10}\selectfont PACE} reproduces the higher energy data points. $^{\text{97}}$Ru could essentially be produced directly from the $^{\text{92}}$Mo($^{\text{7}}$Li, $pn$) ($E_{th}$ = 1.23 MeV), $^{\text{94}}$Mo($^{\text{7}}$Li, $p3n$) ($E_{th}$ = 20.30 MeV) and $^{\text{95}}$Mo($^{\text{7}}$Li, $p4n$) ($E_{th}$ = 28.20 MeV) reactions. Besides, production of $^{\text{97}}$Ru is also observed as the decay from its higher charge isobar, $^{\text{97}}$Rh through $\beta^+$ + EC decay. Since the half-life of the precursor, $^{\text{97}}$Rh (30.7 m), is much smaller than the daughter, $^{\text{97}}$Ru (2.83 d), ($T^p_{1/2}$ $\ll$ $T^d_{1/2}$), independent production of $^{\text{97}}$Ru was calculated following the prescription of Cavinato $et$ $al.$ \cite{cavinato95} :
		\begin{equation}
			\label{eq1}
			\sigma^i = \sigma^c - P^p \bigg[\frac{T_{1/2}^d}{T_{1/2}^d - T_{1/2}^p}\bigg] \sigma^p
		\end{equation} 
	where, 	$\sigma^i$, $\sigma^c$ represent the independent and cumulative cross section of daughter nuclei, respectively; $\sigma^p$ indicates the independent production of precursor radionuclide;\ $P^p$ is the branching ratio of precursor and $T_{1/2}^d$, $T_{1/2}^p$ represent half-lives of daughter and precursor radionuclides, respectively.
	Hence, independent production cross section of $^{\text{97}}$Ru, shown in Fig \ref{fig3}(a), was determined from the equation :

			\begin{equation}
			\label{eq1a}
				         \sigma^i\text{($^{\text{97}}$Ru)} = \sigma^c\text{($^{\text{97}}$Ru)} - 1.0075 \times \sigma^p\text{($^{\text{97}}$Rh)}	 
			\end{equation}
    
 \subsection{$\alpha xn$ channels} 
     
		The measured cross sections of $^{\text{99m,96,95,94,93m+g}}$Tc radionuclides, populated mainly through the $\alpha xn$ channels from the excited composite nucleus are shown in Fig \ref{fig4} and Fig \ref{fig5}(a) and are compared with theoretical estimations. Measured cross section data are significantly large compared to the model calculations obtained from {\fontsize{9}{10}\selectfont PACE} and {\fontsize{9}{10}\selectfont EMPIRE} throughout the observed energy window, except for the $^{\text{94}}$Tc isotope in the low energy range. Contributory reactions for the direct production of these radionuclides are presented in Table \ref{t1}. Cumulative production of $^{\text{93}}$Tc and $^{\text{95}}$Tc arises from  the $^{\text{93m}}$Tc and $^{\text{95}}$Ru radionuclides via IT and EC + $\beta^+$ decay mode, respectively. However, measurement was done immediately after the EOB for the minimum cumulative production.
						
		It is worthy to note that the residues produced through $\alpha$-emitting channels may arise from the CF and/or ICF mechanism. In the CF, $^\text{7}$Li completely fuses with the target and form statistically equilibrated compound nucleus, which may eventually decay via $\alpha xn$ channels. However, in case of a weakly bound projectile like $^\text{7}$Li, it is quite likely that $^\text{7}$Li may break into $\alpha$+triton ($t$) in the nuclear force field before fusion occurs and only one part of the projectile may get fused to the target nucleus forming an incomplete composite system, and the remnant $\alpha$/$t$ moves in the forward cone with approximately the same velocities as projectile. Due to have seven naturally abundant isotopes of $^{\text{nat}}$Mo, various reaction channels contribute to the production of a single residue (Table \ref{t1}). Significantly large cross sections of $^{\text{99m}}$Tc observed in this experiment could be the resultant of the following:
		
	 1. CF of $^{\text{7}}$Li with $^{\text{98}}$Mo leads to production of the $^{\text{99m}}$Tc through $2p4n$ channel
	         	\begin{equation}
	         	  \begin{aligned}
	         	\text{$^\text{7}$Li} + \text{$^{\text{98}}$Mo} \rightarrow \text{[$^{\text{105}}$Rh$^*$]} \rightarrow \text{$^{\text{99m}}$Tc + 2$p$4$n$},\\
	      	         \text{$E_{th}$ = 35.09\ MeV.}
	      	      \end{aligned}   
	         	\end{equation} 
		Other possibilities like $tp2n$, $tdn$ ($d$ represents deuteron) and $2t$ are not shown here, however these channels were included in the theoretical model calculations.
									
	 2. CF of $^{\text{7}}$Li with $^{\text{98}}$Mo leads to production of the $^{\text{99m}}$Tc by $\alpha 2n$ channel
	         	\begin{equation}
	         	   \begin{aligned}
	         		\text{$^{\text{7}}$Li + $^{\text{98}}$Mo $\rightarrow$ [$^{\text{105}}$Rh$^*$] $\rightarrow$ $^{\text{99m}}$Tc + $\alpha$2$n$,}\\
	        	         		\text{$E_{th}$ = 4.77 \ MeV.} 
	               \end{aligned}  
	         	\end{equation}    	
						
	 3. ICF of $t$ with $^{\text{98}}$Mo forming a composite nucleus  $^{\text{101}}$Tc$^*$, which emits two neutron to form $^{\text{99m}}$Tc, and $\alpha$  moves in the forward direction as a spectator.
	 	 \begin{equation}
	 	 \begin{aligned}
	 	 \text{$^\text{7}$Li ($\alpha$ + $t$) $\rightarrow$ $t$ + $^{\text{98}}$Mo  $\rightarrow$ [$^{\text{101}}$Tc$^*$]}  \\ 
	        		\text{$\rightarrow$ $^{\text{99m}}$Tc + 2$n$,}\\
	       	        		\text{$E_{th}$  = 2.04 \ MeV.}
	 	 \end{aligned}  
	 	 \end{equation}  
										
	 4. ICF of $\alpha$ with $^{\text{98}}$Mo forming a composite nucleus  $^{\text{102}}$Ru$^*$, which emits one proton and two neutron (or triton) to form $^{\text{99m}}$Tc, and $t$ moves in the forward direction as a spectator.
	         \begin{equation}
	        	\begin{aligned}
	        		\text{$^\text{7}$Li( $\alpha$ + $t$) $\rightarrow$ $\alpha$ + $^{\text{98}}$Mo  $\rightarrow$ [$^{\text{102}}$Ru$^*$]}  \\ 
	        		\text{$\rightarrow$ $^{\text{99m}}$Tc + $p$ + 2$n$,}\\
	       	        		\text{$E_{th}$  = 22.68 \ MeV.} 
	       	    \end{aligned} 
             \end{equation}            
	
	\subsection{$\alpha pxn$ channels}  
	 Experimentally measured excitation function of $^{\text{93m}}$Mo is compared with the theory as shown in Fig \ref{fig5}(b). {\fontsize{9}{10}\selectfont EMPIRE} satisfactorily reproduces the measured data up to 40 MeV and underestimates beyond it.  However, {\fontsize{9}{10}\selectfont PACE}, which calculates the sum of isomeric and ground state production of $^{\text{93}}$Mo, overestimates the experimental cross section throughout the energy range considered. $^{\text{93m}}$Mo could be produced via $^{\text{92}}$Mo($^{\text{7}}$Li, $\alpha pn$) and $^{\text{94}}$Mo($^{\text{7}}$Li, $\alpha p3n$) reaction channels.  	
	In general, measured data are satisfactorily reproduced by {\fontsize{9}{10}\selectfont EMPIRE} compared to {\fontsize{9}{10}\selectfont PACE} in $xn$-, $pxn$- and $\alpha pxn$-channels. However, the enhancement of cross sections in the $\alpha$-emitting channels provide information about ICF which is discussed in the following subsection.

		\subsection{ICF analysis} 
		The analysis of ICF in $\alpha$-particle emitting channels has been carried out using the data reduction method \cite{gomes04, yadav12} with the theoretical model calculations, {\fontsize{9}{10}\selectfont PACE} and/or {\fontsize{9}{10}\selectfont EMPIRE} which do not include the contribution of ICF process. The enhancement in the measured cross section data over the theoretical evaluation may be regarded as ICF, defined as, say for $\alpha xn$-channel, $\sigma_{ICF} = \Sigma_x\sigma^{expt}_{\alpha xn} - \Sigma_x\sigma^{theor}_{\alpha xn}$, when it explains the other reaction channels with the same set of parameters. Comparative analysis of both the codes with the measured data (Fig \ref{fig2}, \ref{fig3}) reveals that {\fontsize{9}{10}\selectfont EMPIRE} calculation is more reliable compared to {\fontsize{9}{10}\selectfont PACE}. It may be due to accurate treatment of fusion evaluation for heavy-ion induced reactions (CCFUS mechanism), collective (rotational/vibrational) effect energy-dependent level density (EGSM) and consideration of PEQ processes (exciton model) in addition to the compound process. Therefore, {\fontsize{9}{10}\selectfont EMPIRE} calculation was chosen for the analysis of the ICF, which could be obtained from the difference between theoretical and experimental observations. 
		
\begin{figure*} [t]
	  	  	  	\includegraphics[width=180mm]{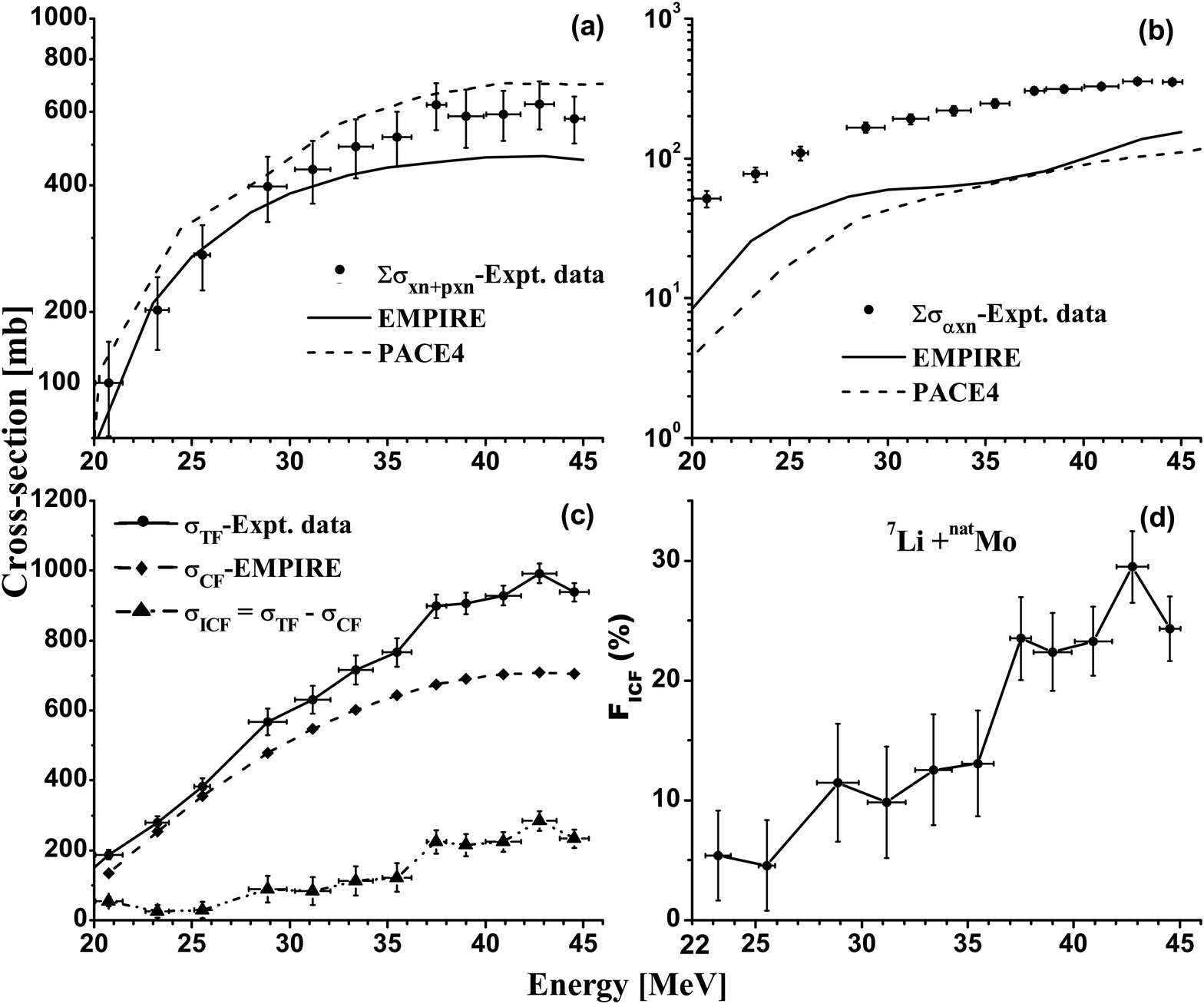}	
	  	  	  	\caption{\label{fig6} Sum of the $xn$, $pxn$ and $\alpha xn$ channels cross sections are compared with the {\fontsize{9}{10}\selectfont EMPIRE} and {\fontsize{9}{10}\selectfont PACE} in 6(a) and 6(b), respectively. Fig 6(c) shows total fusion (TF), complete fusion (CF) and incomplete fusion (ICF) with incident energy. Fig 6(d) represents variation of incomplete fusion fraction with increasing projectile energy.}
	  	   \end{figure*}

		In the Fig \ref{fig6}(a), sum of the measured cross sections of all the $xn$- and $pxn$-channels ($\Sigma\sigma_{xn+pxn}$) are compared with those obtained from {\fontsize{9}{10}\selectfont PACE} and {\fontsize{9}{10}\selectfont EMPIRE}. It is perceived that both the model calculations explain the measured data satisfactory within the observed energy region, which confirms the production of residues via $xn$- and $pxn$-channels through the CF mechanism. A comparison between the sum of the measured cross sections of the $\alpha xn$-emitting channels ($\Sigma\sigma_{\alpha xn}$) and theoretical cross sections estimated with the same set of parameters as used for the other channels is presented in Fig \ref{fig6}(b). Experimental observations were found to be fairly large compared to both the theoretical calculations throughout the energy window, indicating the production of residues through the ICF process in addition to the CF.
		
		Fig \ref{fig6}(c) depicts the sum of cross sections of all the populated residues ($\sigma_{TF}$), sum of the theoretical cross sections of the residues ($\sigma_{CF}$) assessed   from {\fontsize{9}{10}\selectfont EMPIRE}, and the ICF  cross section (i.e., $\sigma_{ICF} = \sigma_{TF} - \sigma_{CF}$). Significant ICF contribution was observed over the CF process. The increasing trend of ICF fraction, F$_{\text{ICF}}$ (F$_{\text{ICF}} = (\Sigma\sigma_{ICF}/\sigma_{TF}^{theor})\times$100, where $\sigma_{TF}^{theor}$ is the total theoretical fusion cross section), with increasing projectile energies was observed in the $^{\text{7}}$Li+$^{\text{nat}}$Mo, as shown in Fig \ref{fig6}(d), similar to that observed in case of the $\alpha$-cluster projectiles \cite{yadav12, sharma14, mukherjee06, hkumar17}. 
		Nevertheless, limitation of the present method includes unobserved $\alpha$-emitting channels that went unidentified due to the short half-lives of the residual radionuclides, or production of stable isotopes; hence the ICF corresponding to these missing channels could not be determined. Thus, the computed ICF cross section ($\sigma_{ICF}$ shown in Fig \ref{fig6}(c)) may be considered as the lower limit of ICF for $^{\text{7}}$Li+$^{\text{nat}}$Mo reaction in the $\alpha$-emitting channels.
		
   	\section{\label{s4}Summary}
		
		This article deals with the measurement of the cross section of evaporation residues produced via CF/ICF in the $^{\text{7}}$Li + $^{\text{nat}}$Mo system within 3--6.5 MeV/nucleon energy region. Experimental cross section data are analysed by comparing the theoretical model calculations, mainly from the Hauser-Feshbach formalism for compound and exciton model for PEQ. The measured excitation functions in the $xn$-, $pxn$-channels are in good agreement with {\fontsize{9}{10}\selectfont EMPIRE} compared to those estimated from {\fontsize{9}{10}\selectfont PACE} that confirms the predominance of the CF process. A relatively good agreement between the experimental data and {\fontsize{9}{10}\selectfont EMPIRE} reveals the dependence of cross sections on the collective discrete states of the interacting nuclei, dependence of rotational/vibrational effect on the level density. However, significant enhancement of cross section in the $\alpha$-emitting channels over the theoretical prediction has been observed and it is attributed to the ICF, which occurred through the breakup of $^{\text{7}}$Li into $\alpha$ + $t$ within the energy range studied. The ICF is found to increase with the increasing projectile energy as reported in the case of several heavy ion reactions induced by the $\alpha$-cluster projectiles. However, measurement of forward angle recoil range distribution of residues and/or spin distributions may further refine ICF in the $^{\text{7}}$Li + $^{\text{nat}}$Mo system.




	 \begin{table*}
	 	\caption{Spectroscopic decay data \cite{nndc} of the measured radionuclides and list of possible contributing reaction channels }
	 	\label{t1}
	 	\begin{ruledtabular}
	 		\begin{center}
	 			\renewcommand{\arraystretch}{0.5}
	 			\begin{tabular}{ccccccc}
	 				Nuclide($J^\pi$) &Decay mode(\%) & Half-life & $E_\gamma$(keV)[$I_\gamma$(\%)] &Reaction &$E_{th}$(MeV)\\
	 				\hline
	 				$^{\text{101m}}$Rh(9/2$^+$) & IT(7.20), EC(92.80) & 4.34 d& 306.86(84.00)  & $^{95}$Mo($^{7}$Li, n)   & 0.0 \\
	 				&       &        &        & $^{96}$Mo($^{7}$Li, $2n$)  & 2.81 \\
	 				&       &        &        & $^{97}$Mo($^{7}$Li, $3n$)  & 10.12 \\
	 				&       &        &        & $^{98}$Mo($^{7}$Li, $4n$)  & 19.38 \\
	 				&       &        &        & $^{100}$Mo($^{7}$Li, $6n$) & 34.57 \\
	 				$^{100}$Rh(1$^{\text{--}}$) & EC + $\beta^+$(100)  &20.8 h & 539.60[80.60]  & $^{95}$Mo($^{7}$Li, $2n$)  &  3.61 \\
	 				&&       & 822.65[21.09]  & $^{96}$Mo($^{7}$Li, $3n$)  & 13.43 \\
	 				&&       & 1553.35[20.67]  & $^{97}$Mo($^{7}$Li, $4n$)  & 20.73 \\
	 				&       &        &        & $^{98}$Mo($^{7}$Li, $5n$)  & 29.98 \\
	 				$^{\text{99m}}$Rh(9/2$^+$) & EC + $\beta^+$($\ge$99.84) & 4.7 h& 340.71[70.00]  & $^{94}$Mo($^{7}$Li, $2n$)  & 4.37 \\
	 				&&       & 617.8[12.00]  & $^{95}$Mo($^{7}$Li, $3n$)  & 12.28 \\
	 				&&       &1261.20[11.00]  & $^{96}$Mo($^{7}$Li, $4n$)  & 22.10 \\
	 				&       &        &        & $^{97}$Mo($^{7}$Li, $5n$)  & 29.40\\
	 				$^{97}$Rh(9/2$^+$)  & EC + $\beta^+$(100)& 30.7 m& 421.55[75.00]  & $^{92}$Mo($^{7}$Li, $2n$)  & 5.86 \\
	 				&       &        &        & $^{94}$Mo($^{7}$Li, $4n$)  & 24.93\\
	 				&       &        &        & $^{95}$Mo($^{7}$Li, $5n$)  & 32.82\\  
	 				$^{97}$Ru(5/2$^+$) & EC + $\beta^+$(100)& 2.83 d& 215.70[85.62] & $^{92}$Mo($^{7}$Li, $pn$)  & 1.23\\
	 				&&       & 324.49[10.79] & $^{94}$Mo($^{7}$Li, $p3n$) & 20.30\\
	 				&       &        &        & $^{95}$Mo($^{7}$Li, $p4n$) & 28.20\\		
	 				$^{95}$Ru(5/2$^+$) & EC + $\beta^+$(100) & 1.64 h& 336.40[70.20]  & $^{92}$Mo($^{7}$Li, $p3n$) & 21.47\\
	 				&&       & 626.63[17.80]  & $^{94}$Mo($^{7}$Li, $p5n$) & 40.5 \\
	 				$^{\text{99m}}$Tc(1/2$^\text{--}$)& IT(99.99) &6.01 h & 140.51[89.00]  & $^{94}$Mo($^{7}$Li, $2p$)  &  0.81\\
	 				&       &        &        & $^{95}$Mo($^{7}$Li, $2pn$) & 8.73 \\
	 				&       &        &        & $^{96}$Mo($^{7}$Li, $\alpha$)   & 0.00\\
	 				&       &        &        & $^{97}$Mo($^{7}$Li, $\alpha n$)  & 0.00 \\
	 				&       &        &        & $^{98}$Mo($^{7}$Li, $\alpha 2n$) & 4.77\\
	 				&       &        &        & $^{100}$Mo($^{7}$Li, $\alpha 4n$)&19.97\\
	 				$^{96}$Tc(7$^+$)  & EC + $\beta^+$(100)& 4.28 d& 778.22[99.76]  & $^{92}$Mo($^{7}$Li, $2pn$)        & 9.40 \\
	 				& &       & 812.54[82.00] & $^{94}$Mo($^{7}$Li, $\alpha n$)  & 0.00\\
	 				&       &        &        & $^{95}$Mo($^{7}$Li, $\alpha 2n$) & 5.96\\
	 				&       &        &        & $^{96}$Mo($^{7}$Li, $\alpha 3n$) & 15.78 \\
	 				&       &        &        & $^{97}$Mo($^{7}$Li, $\alpha 4n$) & 23.08 \\
	 				$^{95}$Tc(9/2$^+$) & EC + $\beta^+$(100) & 20 h  & 765.79[93.8] & $^{92}$Mo($^{7}$Li, $\alpha$)   & 0.00\\
	 				&       &        &        & $^{94}$Mo($^{7}$Li, $\alpha 2n$) & 6.51\\
	 				&       &        &        & $^{95}$Mo($^{7}$Li, $\alpha 3n$) & 14.41\\
	 				$^{94}$Tc(7$^+$) & EC + $\beta^+$(100) & 4.88 h& 702.62[99.60]   & $^{92}$Mo($^{7}$Li, $\alpha n$)  & 0.00 \\
	 				&&       & 871.09[99.99] & $^{94}$Mo($^{7}$Li, $\alpha 3n$) & 17.18 \\
	 				&       &        &        & $^{95}$Mo($^{7}$Li, $\alpha 4n$) & 25.08 \\
	 				$^{\text{93m}}$Tc(1/2$^\text{--}$)&IT(77.4), EC + $\beta^+$(22.6) & 43.5 m& 391.83[58.00] & $^{92}$Mo($^{7}$Li, $\alpha 2n$) & 7.39 \\
	 				&       &        &        & $^{94}$Mo($^{7}$Li, $\alpha 4n$) & 26.45 \\
	 				$^{93}$Tc(9/2$^+$) & EC + $\beta^+$(100) & 2.75 h& 1363.02[66.00]  &  $^{92}$Mo($^{7}$Li, $\alpha 2n$)& 7.39 \\
	 				& &       &1520.37[24.4] & $^{94}$Mo($^{7}$Li, $\alpha 4n$) & 26.45 \\            
	 				$^{\text{93m}}$Mo(21/2$^+$) & IT(99.88), EC + $\beta^+$(0.12)& 6.85 h& 263.06[56.7] & $^{92}$Mo($^{7}$Li, $\alpha pn$) & 3.10\\
	 				& &       & 684.67[99.7] & $^{94}$Mo($^{7}$Li, $\alpha p3n$)& 22.17\\
	 			\end{tabular}
	 		\end{center}
	 	\end{ruledtabular}
	 \end{table*}
	 
	  \begin{table*}
	  	\caption{Cross section (mb) of radioisotopes at various incident energies in the $^{\text{7}}$Li + $^{\text{nat}}$Mo reaction.}
	  	\label{t2}
	  	\begin{ruledtabular}
	  		\begin{center}
	  			\renewcommand{\arraystretch}{1.5}
	  			\begin{tabular}{cccccccc}
	  				\multirow{3}{1.5cm}{\bfseries Energy (MeV)} \\
	  				& \multicolumn{7}{c} { \bfseries Cross-section (mb)}\\
	  				\cline{2-8}
	  				& \multicolumn{1}{c}{$^\text{{101m}}$Rh}
	  				& \multicolumn{1}{c}{$^{\text{100}}$Rh}
	  				& \multicolumn{1}{c}{$^{\text{99m}}$Rh}
	  				& \multicolumn{1}{c}{$^{\text{97}}$Rh}
	  				& \multicolumn{1}{c}{$^{\text{97}}$Ru}
	  				& \multicolumn{1}{c}{$^{\text{95}}$Ru}
	  				& \multicolumn{1}{c}{$^{\text{93m}}$Mo} \\			\hline
	  				20.8 $\pm$ 0.7 & 31.5 $\pm$   6.4 & 47.3 $\pm$  8.6  & 31.1 $\pm$  3.8 & 13.1 $\pm$ 6.0 & 12.7 $\pm$ 3.7 &                 & 0.7 $\pm$ 0.3 \\		
	  				23.2 $\pm$ 0.6 & 42.2 $\pm$   8.4 & 76.4 $\pm$ 12.2  & 51.8 $\pm$  5.7 &  7.1 $\pm$ 4.4 & 24.8 $\pm$ 4.8 &                 & 0.6 $\pm$ 0.4 \\
	  				25.5 $\pm$ 0.4 & 65.4 $\pm$  12.2 & 104.1 $\pm$ 16.1 & 70.4 $\pm$  7.4 &  4.8 $\pm$ 3.5 & 26.2 $\pm$ 4.8 & 1.9 $\pm$ 1.8 & 0.85 $\pm$ 0.3 \\
	  				28.9 $\pm$ 1.0 & 101.3 $\pm$ 13.2 & 155.0 $\pm$ 16.6 & 112.2 $\pm$ 9.0 & 16.2 $\pm$ 8.0 & 9.1 $\pm$ 3.9 & 4.2 $\pm$ 1.6 & 2.2 $\pm$ 0.3 \\
	  				31.2 $\pm$ 0.9 & 123.5 $\pm$ 15.9 & 155.4 $\pm$ 16.7 & 121.2 $\pm$ 9.6 & 10.3 $\pm$ 7.1 & 15.5 $\pm$ 3.0 & 9.9 $\pm$ 2.6 & 3.3 $\pm$ 0.4 \\
	  				33.4 $\pm$ 0.9 & 143.0 $\pm$ 18.3 & 156.9 $\pm$ 16.9 & 132.2 $\pm$ 10.4& 20.3 $\pm$ 9.2 & 14.4 $\pm$ 3.8 &23.5 $\pm$ 4.3 & 4.8 $\pm$ 0.5 \\
	  				35.5 $\pm$ 0.8 & 149.6 $\pm$ 19.1 & 145.2 $\pm$ 15.7 & 137.3 $\pm$ 10.6& 30.5 $\pm$ 11.3& 15.2 $\pm$ 4.9 &34.8 $\pm$ 5.6 & 6.0 $\pm$ 0.6 \\
	  				37.5 $\pm$ 0.5 & 166.9 $\pm$ 21.3 & 130.0 $\pm$  8.5 & 151.7 $\pm$ 11.8& 36.1 $\pm$ 13 & 29.2$\pm$ 6.8 &65.5 $\pm$ 8.9 & 7.8 $\pm$ 0.7 \\
	  				39.0 $\pm$ 0.9 & 169.8 $\pm$ 21.7 & 130.9 $\pm$  8.6 & 149.5 $\pm$ 13.2& 18.9 $\pm$ 5.4 & 59.6 $\pm$ 8.0 &55.7 $\pm$ 7.3 & 8.3 $\pm$ 0.8 \\
	  				40.9 $\pm$ 0.9 & 155.0 $\pm$ 17.2 & 130.1 $\pm$  7.8 & 146.4 $\pm$ 13.7& 19.7 $\pm$ 5.2 & 73.1 $\pm$ 8.0 &67.2 $\pm$ 7.4 & 9.3 $\pm$ 0.8 \\
	  				42.8 $\pm$ 0.8 & 140.4 $\pm$ 15.8 & 144.9 $\pm$  8.6 & 152.7 $\pm$ 14.6& 23.3 $\pm$ 5.4 & 85.7 $\pm$ 9.5 &78.8 $\pm$ 8.3 & 10.3$\pm$ 1.0 \\
	  				44.6 $\pm$ 0.5 & 112.1 $\pm$ 12.8 & 142.8 $\pm$  8.5 & 132.7 $\pm$ 13.0& 23.3 $\pm$ 5.1 & 89.4 $\pm$ 9.7 &77.0 $\pm$ 7.9 & 9.6 $\pm$ 0.9 \\	
	  			\end{tabular}
	  		\end{center}
	  	\end{ruledtabular}
	  \end{table*}

	  \begin{table*}
	  	\caption{Cross section (mb) of radioisotopes at various incident energies in the $^{\text{7}}$Li + $^{\text{nat}}$Mo reaction.}
	  	\label{t3}
	  	\begin{ruledtabular}
	  		\begin{center}
	  			\renewcommand{\arraystretch}{1.5}
	  			\begin{tabular}{ccccccc}
	  				
	  				\multirow{3}{1.5cm}{\bfseries Energy (MeV)} \\
	  				& \multicolumn{6}{c} { \bfseries Cross-section (mb)}\\
	  				\cline{2-7}
	  				& \multicolumn{1}{c}{$^{\text{99m}}$Tc}
	  				& \multicolumn{1}{c}{$^\text{{96}}$Tc}
	  				& \multicolumn{1}{c}{$^{\text{95}}$Tc}
	  				& \multicolumn{1}{c}{$^{\text{94}}$Tc}
	  				& \multicolumn{1}{c}{$^{\text{93m}}$Tc}
	  				& \multicolumn{1}{c}{$^{\text{93}}$Tc}  \\			\hline
	  				20.8 $\pm$ 0.7 & 10.0  $\pm$ 1.5 & 22.0 $\pm$ 2.9  &  4.4 $\pm$ 1.0  & 7.9 $\pm$ 1.0  &                 & 7.2 $\pm$ 1.6  \\		
	  				23.2 $\pm$ 0.6 & 13.2 $\pm$ 2.0 & 31.5 $\pm$ 3.9  & 10.2 $\pm$ 1.8  & 11.1 $\pm$ 1.3 &                 &11.2 $\pm$ 2.0  \\
	  				25.5 $\pm$ 0.4 & 15.8 $\pm$ 2.4 & 45.7 $\pm$ 5.8  & 16.2 $\pm$ 2.6  & 12.7 $\pm$ 1.5 & 1.5 $\pm$ 1.3 &17.2 $\pm$ 2.8  \\
	  				28.9 $\pm$ 1.0 & 22.0 $\pm$ 2.98 & 60.2 $\pm$ 5.0  & 26.8 $\pm$ 2.9  & 14.6 $\pm$ 1.4 & 4.4 $\pm$ 1.5 &38.6 $\pm$ 5.1  \\
	  				31.2 $\pm$ 0.9 & 22.8 $\pm$ 3.1 & 65.4 $\pm$ 5.4  & 36.1 $\pm$ 3.8  & 13.4 $\pm$ 1.3 & 4.6 $\pm$ 2.2 &49.2 $\pm$ 5.9  \\
	  				33.4 $\pm$ 0.9 & 24.0 $\pm$ 3.2 & 70.4 $\pm$ 5.8  & 53.8 $\pm$ 5.3  & 12.9 $\pm$ 1.3 & 5.0 $\pm$ 2.1 &54.1 $\pm$ 6.4  \\
	  				35.5 $\pm$ 0.8 & 23.2 $\pm$ 3.1 & 75.1 $\pm$ 6.2  & 73.5 $\pm$ 6.9  & 12.6 $\pm$ 1.3 & 4.0 $\pm$ 1.7 &59.1 $\pm$ 6.8  \\
	  				37.5 $\pm$ 0.5 & 25.5 $\pm$ 3.4 & 85.2 $\pm$ 7.0  & 108.8 $\pm$ 9.8 & 15.5 $\pm$ 1.5 & 3.6 $\pm$ 1.6 &67.1 $\pm$ 7.6 \\
	  				39.0 $\pm$ 0.9 & 23.0 $\pm$ 3.3 & 89.0 $\pm$ 7.3  & 130.1 $\pm$ 12.2& 18.0 $\pm$ 1.8 & 1.9 $\pm$ 1.0 &51.3 $\pm$ 5.8  \\
	  				40.9 $\pm$ 0.9 & 19.8 $\pm$ 3.3 & 88.8 $\pm$ 6.5  & 150.1 $\pm$ 11.9& 20.0 $\pm$ 2.0 & 0.7 $\pm$ 0.6 &48.1 $\pm$ 5.4  \\
	  				42.8 $\pm$ 0.8 & 19.5 $\pm$ 3.2 & 93.2 $\pm$ 6.9  & 172.5 $\pm$ 13.8& 24.3 $\pm$ 2.5 & 1.0 $\pm$ 0.6 &45.6 $\pm$ 5.2  \\
	  				44.5 $\pm$ 0.5 & 17.6 $\pm$ 2.9 & 85.3 $\pm$ 6.3  & 174 $\pm$ 13.7& 27.5 $\pm$ 2.7 & 0.8 $\pm$ 0.8 &46.1 $\pm$ 5.2  \\
	  			\end{tabular}
	  		\end{center}
	  	\end{ruledtabular}
	  \end{table*}


	\begin{acknowledgments}
		The authors would like to thank the Pelletron staff of BARC-TIFR Pelletron facility for their cooperation and help during the experiment. The work is financially supported by the MHRD (in the form of fellowship),	SINP-DAE-12-plan project TULIP grant, Government of India. 
	\end{acknowledgments}

\end{document}